\documentclass[aps,prd,floatfix,superscriptaddress,amsmath,amssymb,twocolumn,balancelastpage]{revtex4-2}
\pdfoutput=1
\usepackage[dvipsnames]{xcolor}
\usepackage{graphicx}
\usepackage[colorlinks=true,citecolor=blue,linkcolor=blue,breaklinks=true]{hyperref}
\usepackage{natbib}
\usepackage{soul}
\usepackage{gensymb}
\usepackage{mathtools}
\usepackage{multirow}
\usepackage{gensymb}
\usepackage{dutchcal}

\begin{document}

\count\footins = 1000
\title{Supernova-Neutrino-Boosted Dark Matter from All Galaxies}

\author{Yen-Hsun Lin}
\email{yenhsun@phys.ncku.edu.tw}
\affiliation{Institute of Physics, Academia Sinica, Taipei
115, Taiwan}

\author{Meng-Ru Wu}
\email{mwu@gate.sinica.edu.tw}
\affiliation{Institute of Physics, Academia Sinica, Taipei 115, Taiwan}
\affiliation{Institute of Astronomy and Astrophysics, Academia Sinica, Taipei 106, Taiwan}
\affiliation{Physics Division, National Center for Theoretical Sciences, Taipei 106, Taiwan}

\begin{abstract}
It has been recently proposed that the boosted dark matter (BDM) by supernova neutrinos (SN$\nu$) from SN1987a or from the next Galactic supernova (SN) can serve as a novel component to probe nonvanishing interaction between dark matter (DM) and the standard model leptons [Y.-H.~Lin {\it et~al.,} \href{https://doi.org/10.1103/PhysRevLett.130.111002}{Phys.~Rev.~Lett {\bf 130}, 111002 (2023)} and Y.-H.~Lin {\it et~al.,} \href{https://doi.org/10.1103/PhysRevD.108.083013}{Phys.~Rev.~D {\bf 108}, 083013~(2023)}].
In this Letter, we extend this concept and 
evaluate the present-day \emph{diffuse} flux of SN$\nu$~BDM originated from all galaxies at higher redshifts.    
We show that by considering this diffuse BDM (DBDM) component, the best 
sensitivity on the product of the energy-independent DM-$\nu$ and DM-electron cross sections, $\sqrt{\sigma_{\chi\nu}\sigma_{\chi e}}\simeq \mathcal{O}(10^{-37})$~cm$^2$ for sub-MeV DM, can be obtained with large-size neutrino experiments such as Super-Kamiokande or Hyper-Kamiokande, surpassing the estimated SN$\nu$~BDM bound from SN1987a.  
We also examine the impact due to the presence of DM spikes around the supermassive black holes in galaxies on SN$\nu$~BDM and DBDM. 
Our results suggest that both the DBDM and the SN$\nu$~BDM probes are 
robust 
to the uncertain properties of DM spikes, unless the next Galactic SN happens to occur at a location extremely close to or right behind the Galactic Center along the SN line of sight. 
\end{abstract}
\maketitle

\textit{Introduction---}Despite the abundance of compelling evidence for dark matter (DM) in the Universe, its nature remains a puzzling question in fundamental physics. The weakly interacting massive particle was widely perceived as a possible scenario for DM, but its parameter space has been tightly constrained by modern DM direct and indirect  detections \cite{AMS:2015azc,LUX:2016ggv,Fermi-LAT:2017opo,LUX:2017ree,SuperCDMS:2018mne,DAMPE:2017fbg,XENON:2018voc,XENON:2019gfn,XENON:2019zpr,SENSEI:2019ibb,John:2023knt}. Therefore, light DM with mass $m_\chi \lesssim \mathcal{O}({\rm MeV})$ has gained much attention recently, with a plethora of theoretical models being proposed and examined~\cite{Battaglieri:2017aum},  including those allowing for DM interacting with leptons~\cite{He:1991qd,Fox:2008kb,Falkowski:2009yz,Lindner:2010rr,Davoudiasl:2012ag,Chang:2014tea,GonzalezMacias:2015rxl,Battaglieri:2017aum,Chang:2018rso,Foldenauer:2018zrz,Blennow:2019fhy,Escudero:2019gzq,Croon:2020lrf,Lin:2021hen,Workman:2022}. 
 
Boosted DM (BDM), when upscattered by known high-energy cosmic particles, offers a viable method to probe light DM in large underground detectors \cite{Bringmann:2018cvk,Ema:2018bih,Cappiello:2019qsw,Dent:2019krz,Wang:2019jtk,Zhang:2020nis,Guo:2020drq,Ge:2020yuf,Cao:2020bwd,Jho:2020sku,Cho:2020mnc,Lei:2020mii,Guo:2020oum,Xia:2020apm,Dent:2020syp,Ema:2020ulo,Flambaum:2020xxo,Jho:2021rmn,Das:2021lcr,Bell:2021xff,Chao:2021orr,Ghosh:2021vkt,Feng:2021hyz,Wang:2021nbf,Xia:2021vbz,Wang:2021jic,Super-Kamiokande:2022ncz,PandaX-II:2021kai,CDEX:2022fig,Granelli:2022ysi,Cline:2022qld,Xia:2022tid,Cappiello:2022exa,Carenza:2022som,Wang:2023wrx,COSINE-100:2023tcq,Xia:2024ryt}, such as Super-Kamiokande~\cite{Super-Kamiokande:2016yck}, Hyper-Kamiokande~\cite{Hyper-Kamiokande:2018ofw}, DUNE \cite{DUNE:2020ypp} and JUNO \cite{JUNO:2021vlw}. 
In \cite{Lin:2022dbl}, the authors proposed that BDM from supernova neutrinos (SN$\nu$) can provide competitive tests for light DM interacting with known leptons, and  carries time-of-flight  information that  potentially enables a direct measurement of $m_\chi$.  
The follow-up study~\cite{Lin:2023nsm} further explored 
the dependency of this probe on the location of the SN in the Galactic disk,
and applied it to gauged $U(1)_{L_\mu-L_\tau}$ model. 

In this Letter, we further extend this concept by considering the contribution of the SN$\nu$~BDM from galaxies that hosted all past SN explosions 
at different redshifts to the present-day \emph{diffuse} flux. 
Such a diffuse BDM (DBDM) component 
parallels the well known diffuse SN$\nu$ background (DSNB)~\cite{Horiuchi:2008jz,Beacom:2010kk} and thus represents a persistent and isotropic BDM flux. 
We will demonstrate that by analyzing the DBDM signals at large-size neutrino experiments such as Super-Kamiokande (SK) or Hyper-Kamiokande (HK), it can readily offer improved model-independent sensitivities on the product of the DM-$\nu$ and DM-electron cross sections, surpassing the existing constraint from the SN$\nu$~BDM associated with SN1987a. 

In addition to the evaluation of DBDM flux and sensitivity, we also consider the consequence on SN$\nu$~BDM and DBDM due to  
the potential presence of the DM spike~\cite{Gondolo:1999ef,Ullio:2001fb} in the inner halo of a galaxy, resulting from the accretion of DM by the central supermassive black hole (SMBH). 
We will show that for both the SN$\nu$~BDM and DBDM, the associated sensitivities or constraints are not sensitive to the uncertain properties of the DM spike. 
Only when
the next Galactic SN happens to occur extremely close to or right behind the Galactic Center along the SN line of sight, the large DM spike density may substantially enhance the projected 
sensitivity for the case without DM self-annihilation.

\textit{Averaged SN$\nu$~BDM spectrum---}Considering the same 
approach used in Ref.~\cite{Lin:2022dbl} to describe the DM-$\nu$ interaction, 
the energy spectrum of the \emph{total} amount of SN$\nu$~BDM (number per unit energy) for a single SN that explodes at a distance $R$ away from the center of a galaxy is given by 
\begin{equation}\label{eq:BDM_spectrum}
\frac{dN_\chi(R)}{dT_\chi} = (2\pi)^2  \tau
\int d\cos\theta d\cos\theta_c d\ell\ell^2 j_\chi(\ell,\theta,\theta_c,T_\chi),
\end{equation}
where $T_\chi$ is the BDM kinetic energy, $\tau=10$~s is the characteristic SN$\nu$ emission time, 
and $j_\chi$ is the local BDM emissivity, which encodes the information of the DM halo profile in that galaxy and 
the DM--$\nu$ cross section $\sigma_{\chi\nu}$ assumed to be energy-independent and isotropic in center-of-mass frame.
See \cite{SM_refs} for parameter definitions and details. 

Assuming supernovae (SNe) occur near the disk midplane,
we can compute the averaged SN$\nu$~BDM spectrum by averaging over its baryonic mass distribution $\Sigma_b$ projected onto the plane by
\begin{equation}\label{eq:avgBDMspec}
\frac{d\bar N}{dT_\chi} = 
\left .
\int dA \frac{dN_\chi(R)}{dT_\chi} \Sigma_b(R) 
\right /
\int dA\, \Sigma_b(R), 
\end{equation}
where the integrations are performed over the galaxy's differential area $dA$ in the midplane. 

As galaxies of different sizes have different DM halo profiles and baryonic distributions, which affect the exact amount of SN$\nu$~BDM, certain assumptions are needed to model their contribution to DBDM.  
First, we assume that for a galaxy with a known stellar mass $M_G$, its halo mass follows $M_{\rm DM}\simeq \eta M_G$ with $\eta=50$~\cite{Moster:2009fk,Girelli:2020goz}. 
Second, for the halo mass distribution without DM spike, we use the Navarro-Frenk-White (NFW)  profile~\cite{Navarro:1995iw,Navarro:1996gj}, $\rho^{\rm NFW}_\chi$, Eq.~(S13) in \cite{SM_refs}, with $\rho_s=184$~MeV~cm$^{-3}$ and $r_s=24.2$~kpc for a Milky Way (MW) size galaxy and $M_G=M_{\rm MW}=5.29\times 10^{10}$~$M_\odot$. 
For other galaxies, we
take a simple scaling relation $r_s^3\propto M_G$ 
to compute the corresponding $\rho^{\rm NFW}_\chi$. 
The DM profile with  
spike is calculated following \cite{Cline:2023tkp}
For the DM spike density $\rho^{\rm spike}_{\chi}$ that depends on the central SMBH mass and age, we apply the relation
$M_{\rm BH}\approx 7\times10^7 M_\odot \times [M_{\rm DM}/(10^{12}M_\odot)]^{4/3}$~\cite{Ferrarese:2002ct,DiMatteo:2003zx,Baes:2003rt,Cline:2023tkp} for the mass. 
We assume that $t_{\rm BH}$ can be approximated by the age of the Universe at redshift $z$, computed with~\cite{cosmic_calculator} for the DBDM calculation. 
For the SN$\nu$~BDM from MW and SN1987a, we take $t_{\rm BH}\simeq 10$~Gyr. 
Third, for the surface baryonic density $\Sigma_b$, we consider the bulge and the disk components given in \cite{McMillan:2016jtx} for a MW-size galaxy to compute the projected $\Sigma_b^{\rm MW}$. 
For other galaxies, we assume that their bulge-to-disk mass ratios are the same as the MW value, and apply the scaling relations that the scale bulge radius and the scale disk length parameters, $r_{\rm cut}$ and $R_d$ in Eqs. (1) and (3) of \cite{McMillan:2016jtx}, are proportional to the cubic root of the bulge mass and the square root of the disk mass, respectively, to estimate the corresponding $\Sigma_b$. 
Given these assumptions, one can then use Eqs.~\eqref{eq:BDM_spectrum}--\eqref{eq:avgBDMspec} to 
compute $d \bar N/dT_\chi$ for any given $M_G$ of a galaxy.
See \cite{SM_refs} for discussions on DM halo and MW baryonic density profiles.

\textit{DBDM flux---}Knowing the averaged SN$\nu$~BDM energy spectrum from a galaxy, the DBDM flux at redshift $z=0$ consisting of the SN$\nu$~BDM contributions from all galaxies at different redshifts can be written as  
\begin{equation}\label{eq:BDBMflux}
    \frac{d\Phi_\chi}{dT_\chi} = \frac{v_\chi}{H_0} \int_0^{z_{\rm max}} \frac{dz}{\varepsilon(z)}  \int dM_G \frac{d\Gamma_{{\rm SN}}(z)}{dM_G}\frac{d\bar N_\chi(M_G)}{dT_\chi^\prime}, 
\end{equation}
where $H_0=c/(4280\,{\rm Mpc})$ is the Hubble constant, $v_{\chi}=c \sqrt{T_{\chi}(2m_{\chi}+T_{\chi})}/(m_{\chi}+T_{\chi})$ is the BDM velocity,
$\varepsilon(z)=[\Omega_m (1+z)^3+\Omega_\Lambda]^{1/2}$ with $(\Omega_m,\Omega_\Lambda)\approx(0.3,0.7)$ and
$T_\chi^\prime=(1+z)T_\chi$ is the BDM kinetic energy at the source. 
In Eq.~\eqref{eq:BDBMflux}, the term $d\Gamma_{\rm SN}/dM_G$ is the SN rate per comoving volume per galaxy mass at $z$. 
We assume that this rate is approximately proportional to $M_G$ as well as to the known star formation rate per volume $\dot\rho_*(z)$, and can be formulated as 
\begin{equation}
    \frac{d\Gamma_{\rm SN}(z)}{dM_G} = \frac{dn_G(z)}{dM_G} \frac{\dot{\rho}_*(z)}{\dot{\rho}_*(0)} \frac{M_G}{M_{\rm MW}}R_{\rm SN,0}, 
\end{equation}
where $R_{{\rm SN},0}\approx 0.01~{\rm yr}^{-1}$ is the 
SN rate of the MW \cite{Rozwadowska:2020nab} and $\dot{\rho}_*(z)$ is the star formation rate per comoving volume \cite{SM_refs}. 
The term $dn_G(z)/dM_G$ represents the number density of galaxies per comoving volume per galaxy mass at $z$ and
can be parametrized as
\cite{Conselice:2016zid}
\begin{equation}\label{eq:dnG}
    \frac{dn_G(z)}{dm} =\phi_0 \ln10\times 10^{(m-M_c)(1+\gamma)}e^{-10^{m-M_c}},
\end{equation}
where $M_c$ is the characteristic mass in $\log$-10 base, $m=\log_{10}(M_G/M_\odot)$, $\phi_0$ the normalization constant and $\gamma$ the slope for fainter and lower mass galaxies.
They are fitted to observational data at different $z$ \cite{SM_refs}. 

\begin{figure}
\begin{centering}
\includegraphics[width=0.85\columnwidth]{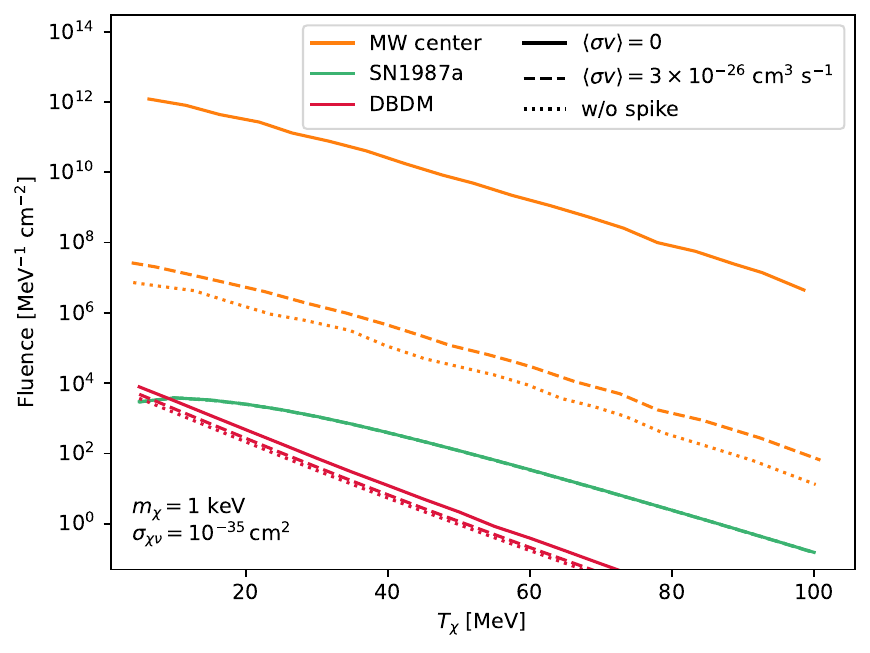}
\end{centering}
\caption{\label{fig:fluence}
The five-year DBDM fluence (red) and the total fluence of SN$\nu$~BDM from SN1987a (green) as well as from the next SN in the MW center (orange) for  
$m_\chi=1$~keV and $\sigma_{\chi\nu}=10^{-35}~{\rm cm^2}$.
The dotted, solid, and dashed lines denote 
different DM profile scenarios.}  
\end{figure}

We compute the DBDM flux given by Eq.~\eqref{eq:BDBMflux} and show in Fig.~\ref{fig:fluence} the corresponding fluence by multiplying the flux with an assumed exposure time $t_{\rm exp}=5$ yr taking $m_\chi=1$~keV and $\sigma_{\chi\nu}=10^{-35}$~cm$^2$ for cases considering the DM spikes with different 
thermally averaged DM self-annihilation cross section $\langle\sigma v\rangle$, as well as the case without DM spikes. 
It shows that the DBDM flux is nearly unaffected by the presence of DM spikes.
This is mainly because the presence of DM spikes only affects a small fraction of SNe that occur very close to the center of galaxies within their spike radii.  

For comparison, we also compute the SN$\nu$~BDM flux using the same set of $m_\chi$ and $\sigma_{\chi\nu}$ for a SN that explodes in the MW center following \cite{Lin:2022dbl} and for SN1987a in the Large Magellanic Cloud (LMC), considering cases with and without spikes. 
Note that here we have taken into account the potentially slight displacement, $\sim 1.75$~kpc, of SN1987a from the LMC center \cite{SM_refs}. 

The resulting SN$\nu$~BDM fluences integrated over the entire duration before the BDM flux vanishes are also shown in Fig.~\ref{fig:fluence} for SN1987a and for the MW center case. 
Without spikes, the 5-yr DBDM fluence can be comparable to the total SN$\nu$~BDM fluence from SN1987a, but several orders of magnitude smaller than that from the next galactic SN at the MW center.   
The energy spectrum of DBDM is somewhat steeper than those from SN$\nu$~BDM, because the DBDM flux is dominated by the contribution at $z\simeq 1-2$, which gets redshifted when arriving at the Earth. 
It is also evident that although the presence of the DM spike without self-annihilation can drastically 
enhance the SN$\nu$~BDM flux if the SN explodes at the center of a galaxy, a small displacement from the center like SN1987a will substantially reduce its effect.
Moreover, a sizable DM annihilation rate  $\langle \sigma v\rangle=3\times 10^{-26}$~cm$^3$~s$^{-1}$ can also largely undermine 
the impact of spike due to the suppressed spike densities.


\begin{figure}
\begin{centering}
\includegraphics[width=\columnwidth]{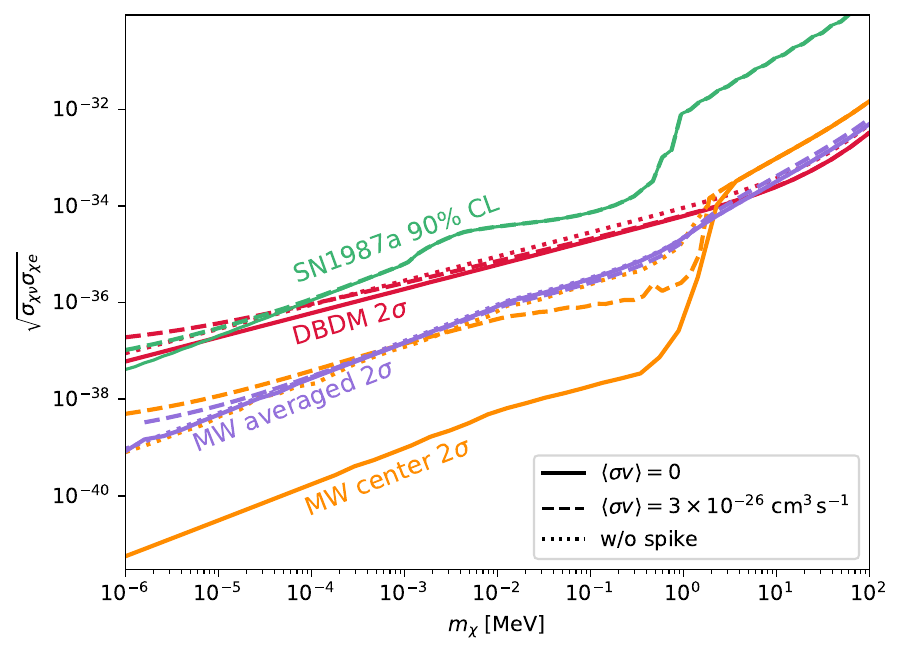}
\end{centering}
\caption{\label{fig:sensitivity-DBDM}
Reevaluated constraint from SN1987a SN$\nu$~BDM (green), the projected sensitivities from DBDM (red), the next galactic SN at MW center (orange), and the averaged projected sensitivity from the next galactic SN (violet) for difference scenarios. 
}
\end{figure}

\textit{Sensitivity---}Within an exposure time $t_{\rm exp}$, the DBDM events $N_{\rm DBDM}$ in a neutrino or DM detector with $N_e$ electron targets for a nonvanishing interaction cross section $\sigma_{\chi e}$ between DM and the electrons can be calculated by 
\begin{equation}
    N_{\rm DBDM} =t_{\rm exp} \times \int_{T_{\rm min}}^{T_{\rm max}} dT_\chi \frac{d\Phi_\chi}{dT_\chi} N_e \sigma_{\chi e}. \label{eq:N_dbdm}
\end{equation}
Taking HK for instance, we set $N_e\approx 7.3\times 10^{34}$, $t_{\rm exp}=5$ yr and $(T_{\rm min},T_{\rm max})=(5,100)$ MeV to estimate $N_{\rm DBDM}$. 
To achieve a $2\sigma$ detection significance, it requires $N_{\rm DBDM}>N_s$ with $N_s$ satisfying $2=N_s/\sqrt{N_s+N_b}$ and $N_b\simeq 5.8\times 10^5$, which 
is the estimated background events in HK in 5~yr also used in \cite{Lin:2023nsm}.
The background events are mainly composed of solar and atmospheric neutrinos, while other radioactive sources are subdominant~\cite{Battistoni:2005pd,Nakano:2017hpu,Dutta:2019oaj}.
For the reevaluated SN1987 bounds, we 
derive the corresponding 90\% confidence level constraint
by utilizing the data from Kamiokande and SK as in Ref.~\cite{Lin:2022dbl}.

We show in Fig.~\ref{fig:sensitivity-DBDM} the projected DBDM sensitivities and the reevaluated SN1987a limits on $\sqrt{\sigma_{\chi\nu}\sigma_{\chi e}}$ as functions of $m_\chi$ for scenarios without DM spikes (red dotted), with DM spikes and $\langle \sigma v \rangle=0$ (red solid), and with DM spikes and $\langle \sigma v \rangle=3\times 10^{-26}$~cm$^3$~s$^{-1}$ (red dashed), respectively. 
Similar to the fluence shown in Fig.~\ref{fig:fluence} for $m_\chi=1$~keV, the DBDM sensitivity and the SN1987a constraint here are nearly unaffected by the property of DM spikes for $m_\chi\gtrsim 0.03$~keV.
For $m_\chi\lesssim 0.03$~keV, self-annihilating DM slightly weakens both limits due to the reduced inner halo densities.
For all scenarios,   
the projected 5-yr DBDM sensitivity from HK can clearly result in better constraint on $\sqrt{\sigma_{\chi\nu}\sigma_{\chi e}}$ than the SN1987a bound for most of the relevant $m_\chi$ range. 
This is mainly because for SN1987a, nearly all or a significant amount of BDM arrives at the Earth within a few years after the explosion when only Kamiokande was operating. 
As a result, the much larger volume of HK can easily provide improved limits on $\sqrt{\sigma_{\chi\nu}\sigma_{\chi e}}$ despite that the DM fluences are comparable (see Fig.~\ref{fig:fluence}).
We also note here that although we do not perform the DBDM analysis using the SK data, we fully expect that such an analysis can yield similar DBDM constraint as our projected 5-yr HK curve, due to the much longer exposure time compensating for the smaller volume.
At $m_\chi\simeq 1$~eV, the SN1987a bound is slightly better than the projected DBDM sensitivity. 
This is because for DBDM, the sensitivity curve exhibits a perfect scaling of $\sqrt{\sigma_{\chi\nu}\sigma_{\chi e}}\propto \sqrt{m_\chi}$ for $m_\chi\lesssim 10$~MeV due to $N_{\rm DBDM}\propto n_\chi\propto 1/m_\chi$.
For the SN1987a curve, it is steeper than $\sqrt{m_\chi}$ for lower $m_\chi$. 
The underlying reason is that for the SN$\nu$~BDM, the duration for nonvanishing flux is proportional to $m_\chi$, given that the SN distance is known~\cite{Lin:2023nsm}.  
For lighter $m_\chi$, it results in lower total background number within the considered exposure time.
As a result, taking a smaller $m_\chi$ gives rise to a better limit than the value inferred from a simple $\sqrt{m_\chi}$ scaling, and therefore leads to steeper dependence of $\sqrt{\sigma_{\chi\nu}\sigma_{\chi e}}$ on $m_\chi$. 

\begin{figure}
\begin{centering}
\includegraphics[width=\columnwidth]{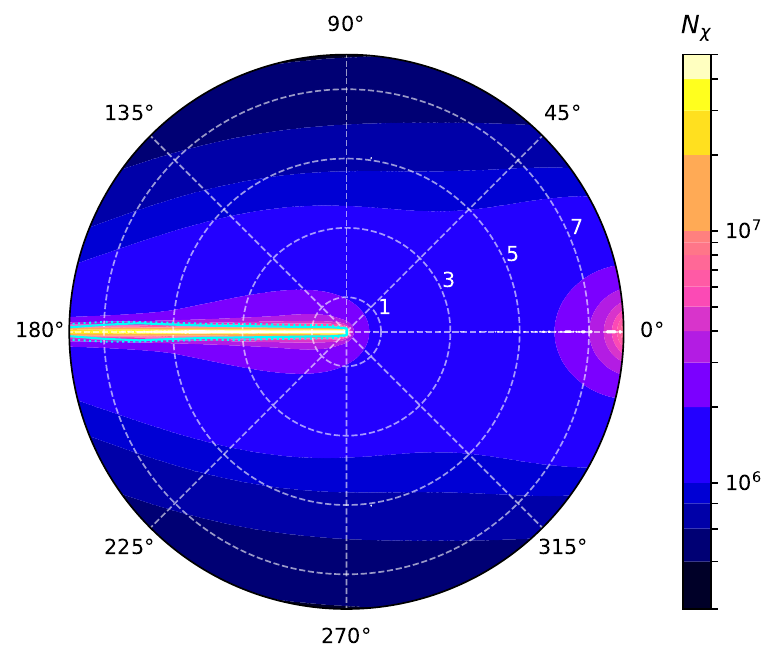}
\end{centering}
\caption{Map of the projected SN$\nu$ BDM events $N_\chi$ at HK for SNe at different locations in the MW disk, taking $m_\chi=1$~keV and $\sqrt{\sigma_{\chi\nu} \sigma_{\chi e}}=10^{-35}~{\rm cm^2}$ for the scenarios considering DM spikes without self-annihilation. 
MW center is at the origin of the coordinates. 
The concentric circles indicate the distances to the MW center in the unit of kpc. 
Earth is located along the $0^\circ$ line outside the map. 
The cyan solid and dotted contours denote the 
event number ratio $N_\chi^{\rm spike}/N_\chi^{\rm w/o}$ between scenarios with and without DM spikes for $N_\chi^{\rm spike}/N_\chi^{\rm w/o}=2$ and 1.5, respectively. 
\label{fig:MWevent}}
\end{figure}

For the projected sensitivities from the next Galactic supernova in MW, we perform two sets of calculations as follows. 
This first set is to assume that the SN is right at the MW center and consider up to 35~yr \footnote{Note again that the duration of having non-vanishing SN$\nu$~BDM flux is typically much less than $\mathcal{O}(1)$~year for $m_\chi\lesssim\mathcal{O}(1)$~keV.} of exposure time with HK. 
Consistent with what shown in Fig.~\ref{fig:fluence}, Fig.~\ref{fig:sensitivity-DBDM} shows that it can provide an improved sensitivity over the SN1987a and DBDM limits by $\sim\mathcal{O}(10-100)$ for the case without DM spike or the scenario of self-annihilating DM with spike.
Without self-annihilation, the presence of DM spike can hugely boost the projected sensitivity by another factor of $\sim\mathcal{O}(100)$. 

However, as inferred from the reevaluated SN1987a cases, the large enhancement from DM spike is not expected for a SN occurring away from the MW center. 
To quantify this, we compute the expected total numbers of SN$\nu$~BDM events for SNe at different locations on the MW disk midplane for the scenario with DM spike and $\langle\sigma v\rangle=0$ as well as that without DM spike. 
Fig.~\ref{fig:MWevent} shows the map of the SN$\nu$~BDM events taking $m_\chi=1$~keV and $\sqrt{\sigma_{\chi\nu} \sigma_{\chi e}}=10^{-35}$~cm$^2$ for the case with DM spike.  
Also shown are the contours delineating constant event ratio of $1.5$ and $2$ between the two scenarios.  
This plot shows clearly that the DM spike only plays a significant role if the next Galactic SN occurs very close to the MW center or right behind the center along the line of sight. 

Given the expected event distribution, we can compute the expected average SN$\nu$~BDM events, by integrating over the occurrence probability 
distribution of Galactic SNe (proportional to MW's baryonic mass surface distribution $\Sigma_b^{\rm MW}$) for $R<8$~kpc. 
For the background estimation, we conservatively take the largest exposure time among all locations for each $m_\chi$, which is defined by the SN location with $R=8$~kpc behind the MW center. 
These allow us to compute the \textit{averaged} projected sensitivities from the next Galactic SN. 
The resulting limits are also shown in Fig.~\ref{fig:sensitivity-DBDM}, which confirm that most likely the very uncertain properties of DM spikes will not affect the expected SN$\nu$~BDM sensitivity from the next Galactic SN. 

\begin{figure}
\begin{centering}
\includegraphics[width=\columnwidth]{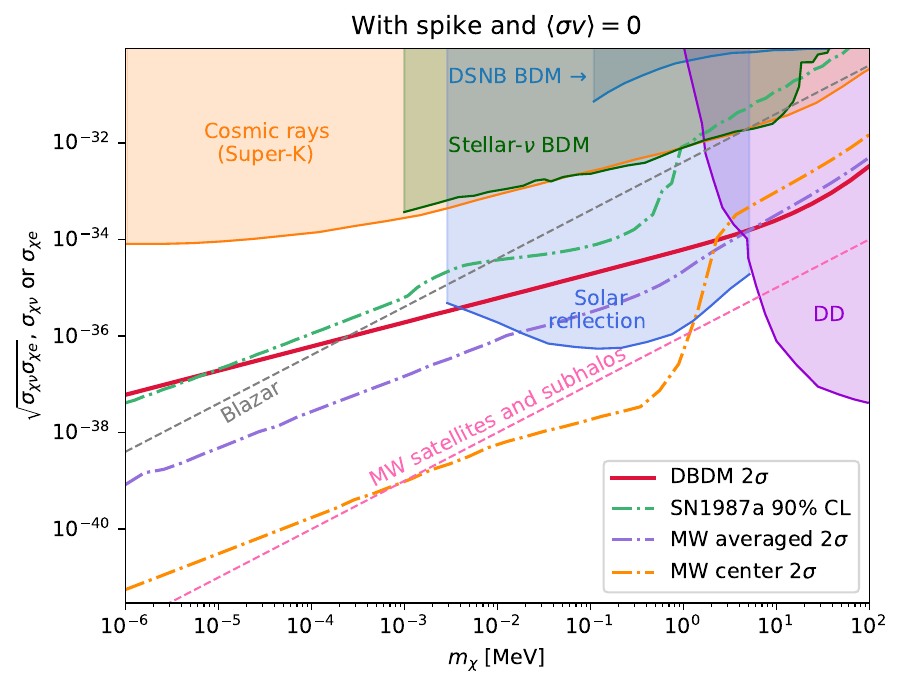}
\end{centering}
\caption{
Comparison of the derived constraint and sensitivities on $\sqrt{\sigma_{\chi\nu}\sigma_{\chi e}}$ from 
this work 
for the scenario considering DM spike without self-annihilation to other existing bounds on $\sqrt{\sigma_{\chi\nu}\sigma_{\chi e}}$~\cite{Jho:2021rmn,DeRomeri:2023ytt}, $\sigma_{\chi\nu}$~\cite{Akita:2023yga,Cline:2023tkp}, and $\sigma_{\chi e}$~\cite{Cappiello:2019qsw,An:2017ojc,XENON:2019zpr,XENON:2019gfn,SENSEI:2019ibb,SuperCDMS:2018mne}.
\label{fig:all}
}
\end{figure}

Fig.~\ref{fig:all} compares the SN1987a limit, the projected sensitivity from 5-yr of DBDM search in HK, and the sensitivity from the next Galactic SN at the MW at center as well as the averaged one derived in earlier sections (with DM spikes and $\langle \sigma v \rangle=0$), to the existing 
bounds on $\sqrt{\sigma_{\chi \nu} \sigma_{\chi e}}$ from the DSNB~BDM~\cite{DeRomeri:2023ytt} and from the stellar-$\nu$~BDM~\cite{Jho:2021rmn}, those on $\sigma_{\chi e}$ from the cosmic-ray BDM~\cite{Cappiello:2019qsw}, solar reflection~\cite{An:2017ojc}, and from direct searches~\cite{XENON:2019zpr,XENON:2019gfn,SENSEI:2019ibb,SuperCDMS:2018mne}, as well as constraints on $\sigma_{\chi\nu}$ from 
blazar~\cite{Cline:2023tkp} and MW satellite and subhalo populations \cite{Akita:2023yga} (see also \cite{Wilkinson:2014ksa,Hooper:2021rjc,Ferrer:2022kei,Brax:2023tvn,Fujiwara:2023lsv,Heston:2024ljf}). 
We note that the latter two categories do not directly constrain $\sqrt{\sigma_{\chi\nu}\sigma_{\chi e}}$ so the naive comparison with our results only makes sense when assuming $\sigma_{\chi\nu}\sim \sigma_{\chi e}$. 
We would also like to caution that all these constraints involve different energy scales in interactions, so the conclusion shown in this plot cannot be directly carried over to scenarios where the cross sections are energy-dependent. 

With the above cautions in mind, we see that both the SN$\nu$~BDM or the DBDM clearly probe $\sqrt{\sigma_{\chi \nu} \sigma_{\chi e}}$ better than the DSNB~BDM and the stellar-$\nu$~BDM \footnote{The authors of \cite{Jho:2021rmn} have shown preliminary results in several workshops on improved constraints by considering the stellar-$\nu$~BDM from earlier galaxies; see e.g.,~\cite{PPC2023}.} by several orders of magnitudes. 
Assuming $\sigma_{\chi\nu}\sim \sigma_{\chi e}$, DBDM and the SN$\nu$~BDM can also provide competitive or even dominating bounds when compared to those that directly probe $\sigma_{\chi e}$ for $m_\chi\lesssim \mathcal{O}(0.1)$~MeV.  
As for the comparison with the existing $\sigma_{\chi \nu}$ bounds, taking the surface values, the SN1987a and the DBDM limit can readily give rise to complementary constraints to the blazar one, and further improvement are likely to be obtained from the next Galactic SN indicated by the MW averaged curve. 
In the most extreme case, if the next Galactic SN occurs very close to the MW center and if the DM spike density can be largely enhanced in a way similar to the profile from taking $\langle \sigma v \rangle=0$, then the SN$\nu$~BDM may independently provide similar constraints on $\sigma_{\chi\nu}$ as that from 
\cite{Cline:2023tkp,Akita:2023yga,Wilkinson:2014ksa,Hooper:2021rjc,Ferrer:2022kei,Brax:2023tvn,Fujiwara:2023lsv,Heston:2024ljf}, 
given $\sigma_{\chi\nu}\sim \sigma_{\chi e}$.

\textit{Summary---}In this Letter, we have further explored the framework of utilizing SN$\nu$~BDM to probe nonvanishing interaction between DM and leptons. 
For the first time, we evaluate the present-day flux of DBDM, which represents the diffuse component of SN$\nu$~BDM from galaxies at all redshifts. 
We have shown that the presence of this intriguing component, conceptually similar to the DSNB, can readily be used to derive the strongest 
bound on $\sqrt{\sigma_{\chi\nu}\sigma_{\chi e}}$. 
Specifically, we have estimated that 
with the consideration of 5-yr exposure time in a neutrino experiment similar to HK, the resulting sensitivity will be better than the SN$\nu$~BDM constraint based on SN1987a from \cite{Lin:2022dbl}, probing  $\sqrt{\sigma_{\chi\nu}\sigma_{\chi e}}$ to the level of $\mathcal{O}(10^{-36})$~cm$^2$ for $m_\chi\lesssim \mathcal{O}(1)$~MeV. 
This result implies that a reanalysis of SK data accumulated over the past three decades could allow to place a similarly leading limit on $\sqrt{\sigma_{\chi\nu}\sigma_{\chi e}}$.

We have also considered the impact due to the presence of DM spikes around the SMBHs in galaxies on DBDM and SN$\nu$~BDM from SN1987a and from the next Galactic SN.  
For the SN1987a SN$\nu$~BDM, we reevaluated the constraint by taking into account the potentially small displacement of SN1987a from the LMC center.  
Taking three specific scenarios of the spike profiles corresponding to cases with and without DM self-annihilation, we have shown that both the DBDM sensitivity and the reevaluated SN1987a bound are insensitive
to the uncertain profile of DM spikes.
For the next Galactic SN, we have observed that the presence of spike can significantly enhance the SN$\nu$~BDM flux and the associated sensitivity only if the SN occurs very close to the MW center or right behind the center along the line of sight. 
For most regions on the MW disk where the SNs can occur, the resulting SN$\nu$~BDM sensitivity is also insensitive to the presence of spike.
Given these, the next Galactic SN will likely offer further improved sensitivity on $\sqrt{\sigma_{\chi\nu}\sigma_{\chi e}}$ by another factor of 10--100 over the SN1987a or the DBDM limits.  

Our results highlight the rich phenomena and the significant discovery potential associated with SN$\nu$~BDM. 
We expect that further improved limits and bounds on specific particle physics models can be deduced, similarly to what demonstrated in \cite{Lin:2023nsm}. 
It will also be worthy to incorporate DM nucleon interaction into this framework. We leave such explorations to future work.  

\bigskip
\textit{Acknowledgments---}We are grateful to useful discussions with Yongsoo Jho and Jong-Chul Park and valuable comments from Paolo Salucci. 
Y.-H. L. and M.-R.~W.~acknowledge supports from the National Science and Technology Council, Taiwan under Grant No.~111-2628-M-001-003-MY4, the Academia Sinica under Project No.~AS-CDA-109-M11, and Physics Division, National Center for Theoretical Sciences of Taiwan. 
We would like to also acknowledge the use of computational resources provided by the Academia Sinica Grid-computing Center (ASGC). 

\bibliographystyle{apsrev4-1}
\bibliography{main}

\clearpage
\newpage
\maketitle
\onecolumngrid

\setcounter{equation}{0}
\setcounter{figure}{0}
\setcounter{section}{0}
\setcounter{table}{0}
\setcounter{page}{1}
\makeatletter

\renewcommand{\theequation}{S\arabic{equation}}
\renewcommand{\thefigure}{S\arabic{figure}}
\renewcommand{\thetable}{S\arabic{table}} 

\begin{center}
\textbf{\large Supernova-neutrino-boosted dark matter from all galaxies}

\vspace{0.05in}
{ \it \large Supplemental Material}\\ 
\vspace{0.05in}
{Yen-Hsun Lin and Meng-Ru Wu}
\end{center}

\section{The Scheme in individual galaxy}

\subsection{The boosted dark matter emissivity}

\begin{figure}[t]
\begin{centering}
\includegraphics[width=0.35\columnwidth]{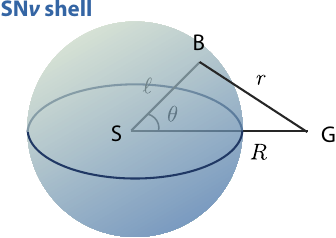}
\end{centering}
\caption{\label{fig:scheme}
Schematic plot of DM boosted by SN$\nu$ within an expanding spherical shell. The SN occurs at $\mathsf{S}$, the galactic center is located at
$\mathsf{G}$ while DM is boosted at $\mathsf{B}$.
}
\end{figure}

The scheme of supernova-neutrino-boosted dark matter (SN$\nu$ BDM) from an individual galaxy is shown in Fig.~\ref{fig:scheme}. Three locations $\mathsf{S}$, $\mathsf{G}$ and $\mathsf{B}$ refer to SN, galactic center (GC) and scattering point respectively. Thus, the local BDM emissivity $j_\chi$ in Eq.~(1) at $\mathsf{B}$ is, following the definition in \cite{Lin:2022dbl},
\begin{equation}\label{eq:emissi}
j_{\chi}(\ell,\theta,\theta_c,T_{\chi})= c n_{\chi}(r(\ell))
\frac{d\sigma_{\chi\nu}}{d\Omega_c}
\frac{dn_{\nu}}{dE_{\nu}}
\left(\frac{dE_{\nu}}{dT_{\chi}} \frac{v_\chi}{c}\right), 
\end{equation}
where \
$r^2(\ell)=\ell^2+R^2-2\ell R \cos\theta$ is the distance from $\mathsf{B}$ to GC,
$\theta_c$ is the scattering angle of the BDM in the center-of-mass (CM) frame and $v_{\chi}/c=\sqrt{T_{\chi}(2m_{\chi}+T_{\chi})}/(m_{\chi}+T_{\chi})$ is the BDM velocity.
In Eq.~\eqref{eq:emissi}, $n_\chi(r)=\rho_\chi(r)/m_\chi$ is the DM number density at $\mathsf{B}$
with $\rho_\chi(r)$ the DM mass density (discussed in the next section), and the differential DM--$\nu$ interaction cross section in the CM frame $d\sigma_{\chi\nu}/d\Omega_c=\sigma_{\chi\nu}/4\pi$ is assumed to be isotropic in the CM frame where $d\Omega_c$ is the differential solid angle. 
For the neutrino number density within the SN$\nu$ shell,
$dn_{\nu}/dE_{\nu}$, we use the same form as in \cite{Lin:2022dbl,Lin:2023nsm}, 
\begin{equation}
\frac{dn_{\nu}}{dE_{\nu}} =\sum_{i}\frac{L_{\nu_{i}}}{4\pi \ell^{2}\langle E_{\nu_{i}}\rangle}E_{\nu}^{2}f_{\nu_{i}}(E_{\nu}), \label{eq:SNnu_spectrum}
\end{equation}
where $L_{\nu_{i}}=L_{\nu,{\rm tot}}/6$ 
is the luminosity 
of each flavor ($\nu_e$,~$\nu_\mu$,~$\nu_\tau$ and their antineutrinos).  
The average energy $\langle E_{\nu_e}\rangle$,~$\langle E_{\bar\nu_e}\rangle$, 
and~$\langle E_{\nu_x}\rangle$ ($\nu_x\in \{ \nu_\mu, \nu_\tau, \bar\nu_\mu, \bar\nu_\tau\}$) are taken to be 11,~16,~25~MeV, respectively \cite{Duan:2006an}.  
We assume a Fermi-Dirac distribution $f_{\nu_{i}}$ with a pinch parameter $\eta_{\nu_i}\equiv\mu_{\nu_i}/T_{\nu_i}=3$, such that $T_{\nu_i} \approx \langle E_{\nu_i}\rangle/3.99$  \cite{Duan:2006an}. 
The term $dE_\nu/dT_\chi$ in Eq.~\eqref{eq:emissi} is the Jacobian factor that converts $dE_\nu$ to $dT_\chi$,
\begin{equation}
\frac{dE_\nu}{dT_\chi} =
\frac{1}{2c_2}\left(1+\frac{1+c_2 m_\chi/T_\chi}{\sqrt{2c_2 m_\chi/T_\chi + 1}}\right), 
\end{equation}
with $c_2=\cos^2(\theta_c/2)$.

\subsection{Milky Way baryon density profile}

We follow \cite{McMillan:2016jtx} to model the Milky Way (MW) stellar density distribution by considering the bulge component $\rho_b$,
\begin{equation}
    \rho_b(R,z)=\frac{\rho_{0,b}}{[1+r^\prime(R,z)/r_0]^\alpha}\exp\left[-\frac{r^{\prime 2}(R,z)}{r_{\rm cut}^2}\right]
\end{equation}
and the stellar disc component $\rho_d$,
\begin{equation}
    \rho_d(R,z)=\frac{\Sigma_0}{2z_d}\exp\left(-\frac{|z|}{z_d}-\frac{R}{R_d}\right),
\end{equation}
where $R$ and $z$ are the radial distance and height to MW center in cylindrical coordinate.
For the bulge, $r^\prime = \sqrt{R^2+(z/q)^2}$ with $\alpha=1.8$, $r_0=0.075$ kpc, $r_{\rm cut}=2.1$ kpc, $q=0.5$ and $\rho_{0,b}=9.93\times 10^{10}M_\odot\,{\rm kpc}^{-3}$.
For the stellar disc component, it can be further divided into thin and thick ones. The corresponding parameters are listed in Table~\ref{tab:stellar_disc}.
We ignore the contribution from gas as it is insignificant when estimating the average SN position.

To obtain the MW area density $\Sigma_b^{\rm MW}(R)$, we integrate over the vertical height $z$ by
\begin{equation}
    \Sigma^{\rm MW}_b(R) = 2\int_0^{z_{\rm max}} dz~[\rho_b(R,z)+\rho_d(R,z)]
\end{equation}
where factor $2$ accounts for symmetric $\rho_{b,d}(R,z)$ under $z\to-z$ and $\Sigma_b^{\rm MW}(R)$ is insensitive to the choice of $z$ when $z\gg z_{\rm max}$.
In this work, we have $z_{\rm max}=10$ kpc.

\section{DM density profile with spike}\label{appx:DM_profile}

\begin{table}
\begin{centering}
\begin{tabular}{|c|c|c|c|}
\hline 
 & $\Sigma_{0}$ ($M_{\odot}\,{\rm kpc}^{-3}$) & $R_{d}$ (kpc) & $z_{d}$ (kpc)\tabularnewline
\hline 
\hline 
thin & $8.96\times10^{8}$ & 2.5 & 0.3\tabularnewline
\hline 
thick & $1.83\times10^{8}$ & 3.02 & 0.9\tabularnewline
\hline 
\end{tabular}
\par\end{centering}
\caption{Values for parameters describing stellar dics.\label{tab:stellar_disc}}
\end{table}

The DM density in the presence of a spike in a galaxy has been extensively exploited in \cite{Gondolo:1999ef,Ullio:2001fb,Cline:2023tkp} and can be conveniently parametrized by~\cite{Cline:2022qld,Cline:2023tkp} 
\begin{equation}
    \rho_\chi^{\rm spike}(r)=
    \begin{dcases}
        0, & r < 4R_S, \\
        \frac{\rho_{\alpha}(r)\rho_{c}}{\rho_{\alpha}(r)+\rho_{c}}, & 4R_S \leq r < R_{\rm spike},\\
        \frac{\rho_{\chi}^{\rm NFW}(r)\rho_c}{\rho_{\chi}^{\rm NFW}(r)+\rho_c}, & r \geq R_{\rm spike},
    \end{dcases}\label{eq:spike_profile}
\end{equation}
where $r$ is the radius from the galactic center, 
$\rho_c=m_\chi/(\langle \sigma v\rangle t_{\rm BH})$ is the saturation density considering the effect of DM self-annihilation, $\langle \sigma v\rangle$ denotes the thermally averaged DM annihilation cross section,  
$R_S=2GM_{\rm BH}/c^2$ is the Schwarzschild radius of the central SMBH,
$M_{\rm BH}$ is the SMBH mass, and $t_{\rm BH}$ is the SMBH age. 
The function $\rho_\alpha$ denotes the spike profile inside the radius $R_{\rm spike}$ without DM self-annihilation. 
For $\rho_\alpha$, the index parameter  
$\alpha$ determines the slope of the spike profile in the inner halo. 
In \cite{Cline:2022qld,Cline:2023tkp}, two values of $\alpha = 7/3$ and $3/2$ were assumed. 
$\alpha = 7/3$ represents the original value proposed in \cite{Gondolo:1999ef,Ullio:2001fb}.
However, the lower dynamical relaxation resulting from gravitational scattering between DM and stars may reduce $\alpha$ to $3/2$, leading to a less cuspy profile~\cite{Gnedin:2003rj}. 
We conservatively take $\alpha = 3/2$ in this work. 
The explicit expressions of $\rho_{3/2}$ and the corresponding $R_{\rm spike}$ are given by~\cite{Cline:2023tkp}

\begin{figure}[t]
\begin{centering}
\includegraphics[width=0.4\columnwidth]{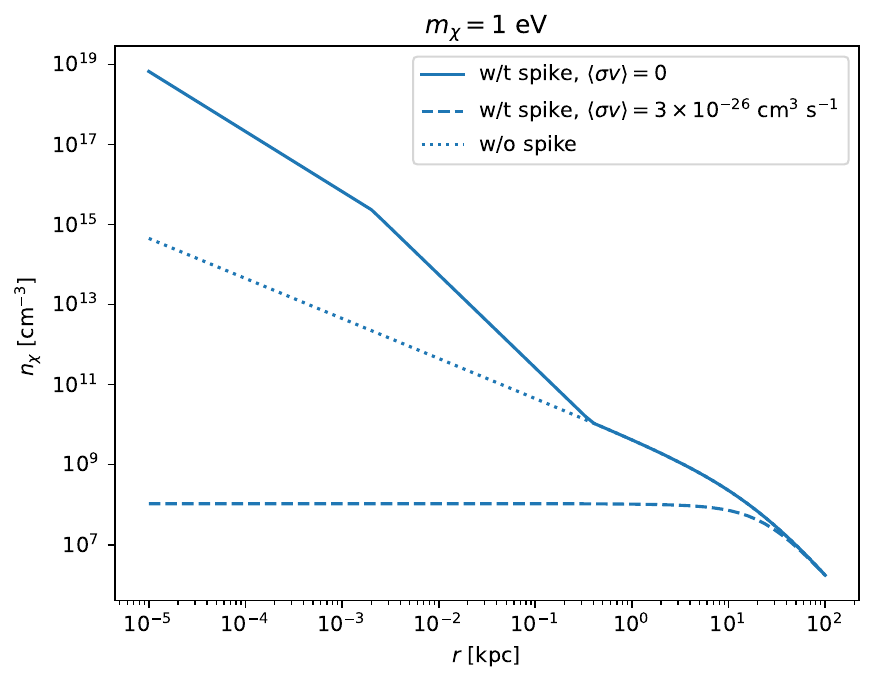}\quad
\includegraphics[width=0.4\columnwidth]{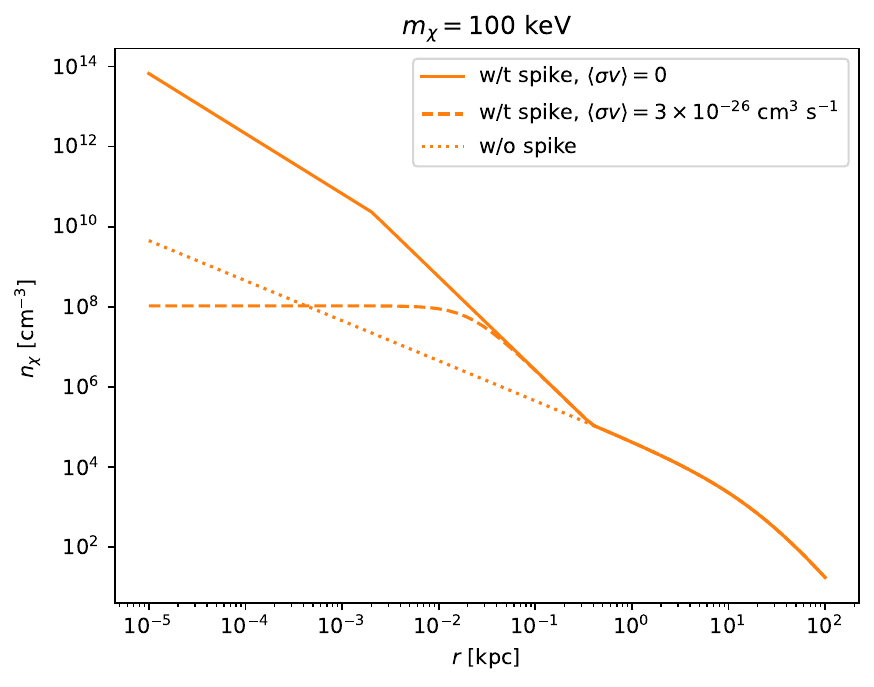}
\end{centering}
\caption{\label{fig:DM_profiles}
MW DM profile for $m_\chi=1$~eV (left) and $100$~keV (right).
The solid lines indicate DM profiles containing spikes for scenario without annihilation,
the dashed lines for DM profiles including spikes but with $\langle \sigma v\rangle=3\times 10^{-26}$~cm$^3$~s$^{-1}$, and the 
dotted lines the original NFW profiles.} 
\end{figure}

\begin{equation}\label{eq:rho32}
\rho_{3/2}(r)\simeq\begin{dcases}
\rho_{N}\left(1-\frac{r_{i}}{r}\right)^{3}\left(\frac{r_{h}}{r}\right)^{3/2}, & r_{i}\leq r<r_{h}\\
\rho_{N}^\prime\left(\frac{R_{\rm spike}}{r}\right)^{7/3}, & r>r_{h}
\end{dcases}
\end{equation}
where $\rho_{N}=\mathcal{N}r_h^{-3/2}$, $\rho_{N}^\prime=\rho_N(r_h/R_{\rm spike})^{7/3}$, $r_i=4R_S$ and $r_h=GM_{\rm BH}/\sigma_\star^2$ is the SMBH influence radius.  
The stellar velocity dispersion $\sigma_\star$ in the galaxy bulge can be
estimated by \cite{McConnell:2011mu}
\begin{equation}
    \log_{10}\left(\frac{M_{\rm BH}}{M_\odot}\right)=8.29 +5.12\log_{10}\left(\frac{\sigma_\star}{200\,{\rm km\,s^{-1}}}\right).
\end{equation}
The spike radius $R_{\rm spike}$ is given by 
\begin{equation}
    R_{\rm spike}=\left(\frac{\mathcal{N}}{\rho_{s}r_{s}}\right)^{3/4}r_{h}^{5/8}, 
\end{equation}
with the normalization constant
\begin{equation}
\mathcal{N}=\frac{M_{\rm BH}}{4\pi[f_{3/2}(r_h)-f_{3/2}(r_i)]}, 
\end{equation}
where
\begin{equation}
f_\alpha(r)\equiv  r^{-\alpha}\left(\frac{r^3}{3-\alpha}+\frac{3r_ir^2}{\alpha-2}-\frac{3r_i^2 r}{\alpha-1}+\frac{r_i^3}{\alpha}\right).
\end{equation}
Without DM self-annihilation, $\rho_{\chi}^{\rm spike}(r)$ relaxes to 
the conventional Navarro-Frenk-White profile~\cite{Navarro:1995iw,Navarro:1996gj} at $r>R_{\rm spike}$, 
\begin{equation}\label{eq:NFWprofile}
    \rho_\chi^{\rm NFW}(r)=\frac{\rho_s}{\frac{r}{r_s}(1+ \frac{r}{r_s})^{2}},
\end{equation}
where $r_s$ and $\rho_s$ characterize the scale radius and the density of the halo. 
For self-annihilating DM, a nonzero $\langle \sigma v\rangle$ reduces the spike density and produces a relatively cored profile that saturates at $\rho_c$.


Numerical results of MW DM number density $n_\chi(r)=\rho^{\rm spike}_\chi(r)/m_\chi$ for $m_\chi=1\,$eV (left) and $100\,$keV (right)
are shown in Fig.~\ref{fig:DM_profiles}.
Without annihilation, the spike profile simply takes the form given by Eq.~\eqref{eq:rho32}, which is independent of $m_\chi$ in $r<R_{\rm spike}$.  
When taking non-zero $\langle \sigma v\rangle$, the DM self-annihilation can significantly reduce the central DM density, giving rise to cored profiles saturate at $\rho_c$.  
For smaller $m_\chi$, the annihilation effect can also affect the radial range outside $R_{\rm spike}$, illustrated by the left panel of the figure.  

We note that even without a spike, the DM density in the inner halo region could be shallower than the NFW profile, as discussed in \cite{Salucci:2018hqu}. 
However, we expect that when taking into account the average over area density for the case of DBDM or the averaged MW SN$\nu$ sensitivity, the exact shape of the inner halo has minimal impact on our result. 

\section{$\dot{\rho}_*(z)$ and fitted values for $dn_G/dm$ }

We take the star formation rate per comoving $\dot\rho_*(z)$ in Eq.~(6) by
\begin{equation}
    \dot{\rho}_*(z)=\dot{\rho}_0\left[(1+z)^{a\eta}+\left(\frac{1+z}{B}\right)^{b\eta}+\left(\frac{1+z}{C}\right)^{c\eta}\right]^{1/\eta}
\end{equation}
from \cite{Yuksel:2008cu}, with
$\dot{\rho}_0=0.0178M_\odot~{\rm yr}^{-1}~{\rm Mpc}^{-3}$, $(a,b,c,\eta)=(3.4,-0.3,-3.5,-10)$, and the redshift break constants $B=(1+z_1)^{1-a/b}$ and $C=(1+z_1)^{(b-a)/c}(1+z_2)^{1-b/c}$ at $z_1=1$ and $z_2=4$, respectively.
For the values of parameters used in $dn_G/dm$, Eq.~(7) of the main text, they are documented in the Table I of  \cite{Conselice:2016zid} and listed explicitly in Table~\ref{tab:dnG_params}.

\begin{table}[t]
\begin{centering}
\begin{tabular}{|c|c|c|c|c|}
\hline 
Redshift ($z$) & $\gamma$ & ${\rm log}_{10}(M_{c}/M_{\odot})$ & $\phi_{0}\, (\times10^{-4})$ (Mpc$^{-3}$) & References\tabularnewline
\hline 
\hline 
$0.0-0.7$ & $-1.11$ & $11.22$ & $18.2$ & \multirow{2}{*}{Fontana+04 \cite{Fontana:2004jy} in \cite{Conselice:2016zid}}\tabularnewline
\cline{1-4} \cline{2-4} \cline{3-4} \cline{4-4} 
$0.7-1.0$ & $-1.27$ & $11.37$ & $11.0$ & \tabularnewline
\hline 
$1.0-1.4$ & $-1.28$ & $11.26$ & $6.2$ & \multirow{7}{*}{Fontana+06 \cite{Fontana:2006xg} in \cite{Conselice:2016zid}}\tabularnewline
\cline{1-4} \cline{2-4} \cline{3-4} \cline{4-4} 
$1.4-1.7$ & $-1.31$ & $11.25$ & $4.3$ & \tabularnewline
\cline{1-4} \cline{2-4} \cline{3-4} \cline{4-4} 
$1.8-2.2$ & $-1.34$ & $11.22$ & $3.1$ & \tabularnewline
\cline{1-4} \cline{2-4} \cline{3-4} \cline{4-4} 
$2.2-2.6$ & $-1.38$ & $11.16$ & $2.4$ & \tabularnewline
\cline{1-4} \cline{2-4} \cline{3-4} \cline{4-4} 
$2.6-3.0$ & $-1.41$ & $11.09$ & $1.9$ & \tabularnewline
\cline{1-4} \cline{2-4} \cline{3-4} \cline{4-4} 
$3.0-3.5$ & $-1.45$ & $10.97$ & $1.5$ & \tabularnewline
\cline{1-4} \cline{2-4} \cline{3-4} \cline{4-4} 
$3.5-4.0$ & $-1.49$ & $10.81$ & $1.1$ & \tabularnewline
\hline 
$4.0-4.5$ & $-1.53$ & $10.44$ & $3.0$ & \multirow{4}{*}{Song+16 \cite{Song_2016} in \cite{Conselice:2016zid}}\tabularnewline
\cline{1-4} \cline{2-4} \cline{3-4} \cline{4-4} 
$4.5-5.5$ & $-1.67$ & $10.47$ & $1.3$ & \tabularnewline
\cline{1-4} \cline{2-4} \cline{3-4} \cline{4-4} 
$5.5-6.5$ & $-1.93$ & $10.30$ & $0.3$ & \tabularnewline
\cline{1-4} \cline{2-4} \cline{3-4} \cline{4-4} 
$6.5-8.0$ & $-2.05$ & $10.42$ & $0.1$ & \tabularnewline
\hline 
\end{tabular}
\par\end{centering}
\caption{Fitted values for parameters in Eq.~(7) of the main text.\label{tab:dnG_params}}

\end{table}


\section{SN1987a and LMC}


To estimate the SN$\nu$ BDM coming from SN1987a, we acquire more accurate positioning data from \cite{nasa_ned}.
For SN1987a, it occurs at 51.4~kpc~\cite{Panagia:2003rt} away with right ascension (RA) of $05^h\,35^m\,27.8733^s$ and declination (DEC) of $-69^\circ\,16'\,10.478''$~\cite{nasa_ned}. 
The center of LMC was estimated to be around 49.97~kpc away~\cite{Pietrzynski:2013gia}, with RA of $05^h\,23^m\,34.5264^s,$ and DEC of $-69^\circ \,45'\,22.053''$ \cite{nasa_ned}.
The displacement between the two objects is approximate 1.75~kpc, however, the associated uncertainties are relatively large so that the SN1988a can be much closer to the LMC center.
Furthermore, the LMC's spike profile is calculated using Eq.~\eqref{eq:spike_profile} by taking $M_{\rm BH}=10^7 M_\odot$ \cite{Boyce_2017} and replacing $\rho_\chi^{\rm NFW}$ with the Hernquist halo profile \cite{Lin:2022dbl,Erkal_2019}
\begin{equation}
    \rho_\chi^{\rm Hernquist}(r)=\frac{\rho_s}{\frac{r}{r_s}(1+ \frac{r}{r_s})^{3}},
\end{equation}
where $r_s=31.9$~kpc and $\rho_s=68~{\rm MeV\,cm^{-3}}$. 

\section{Code availability}

We provide a python package \texttt{dukes} on PyPI that can fully reproduce the DBDM results in this work. See its project page  \cite{dukes2024} for installation and usage. 

\end{document}